\documentstyle[prb,aps]{revtex}

\begin{document}

\draft

\preprint{\today}

\title{Electronic Structure and Charge Dynamics of Huesler Alloy
Fe$_2$TiSn Probed by Infrared and Optical Spectroscopy}

\author{S.V.~Dordevic and D.N.~Basov}
\address{Department of Physics, University of California,
San Diego, La Jolla, CA 92093}

\author{A.~\'{S}lebarski$^{**}$ and M.B.~Maple}
\address{Department of Physics and
Institute for Pure and Applied Physical Sciences, \\ University of
California, San Diego, La Jolla, CA 92093}

\author{L.~Degiorgi}
\address{Laboratorium f$\ddot{u}$r Festk$\ddot{o}$rperphysik
ETH-Z$\ddot{u}$rich, 8093 Z$\ddot{u}$rich, Switzerland}

\wideabs{

\maketitle

\begin{abstract}
We report on the electrodynamics of a Heusler alloy Fe$_2$TiSn
probed over four decades in energy: from the far infrared to the
ultraviolet. Our results do not support the suggestion of
Kondo-lattice behavior inferred from specific heat measurements.
Instead, we find a conventional Drude-like response of free
carriers, with two additional absorption bands centered at around
0.1 and 0.87 eV. The latter feature can be interpreted as
excitations across a pseudogap, in accord with band structure
calculations.
\end{abstract}
}

\narrowtext

Fe$_2$TiSn belongs to a large group of materials commonly referred
to as Heusler alloys with the general formula X$_2$YZ, where X and
Y are transition metals and Z is a non-magnetic element. Hesuler
and closely related half-Heusler (with the formula XYZ) alloys
have been the subject of continued interest for almost 70 years
\cite{heusler34,fujita72,mancoff99,degiorgi02}. A member of this
family Fe$_2$VAl has recently attracted a lot of attention in
connection with possible $d$-electron heavy fermion (HF) behavior
\cite{nishino97}. The resistivity of Fe$_2$VAl displayed anomalous
temperature dependence and specific heat measurements revealed an
upturn in $C_v(T)$ resembling that of conventional $f$-electron HF
compounds \cite{nishino97}. However when the specific heat
measurements were repeated in high magnetic field they showed that
the up-turn was due to a Schotky anomaly arising from magnetic
clusters, not the Kondo interaction \cite{lue99}. A number of band
structure calculations yield only a minor mass renormalization
\cite{weht98,singh98,weinert98,guo98}. An infrared (IR) study
reported recently for Fe$_2$VAl (Ref.~\onlinecite{okamura00}) also
finds no characteristic features of the HF state in the
electrodynamic response of this compound.

Based on electrical resistivity and specific heat measurements
Fe$_2$TiSn has also been speculated to be a HF metal with the
quasiparticle effective mass of $\sim$ 40 m$_e$, where m$_e$ is
the free electron mass \cite{slebarski00,slebarski01}. Note
however that the electronic contribution $\gamma$ to the specific
heat $C_v(T)=\gamma T + \beta T^3$ is relatively small (12 mJ
mol$^{-1}$ K$^{-2}$) compared with conventional HF metals.
Moreover, a similar temperature dependence of $C_v(T)$ in
Fe$_2$VAl had first been proposed to be Kondo in origin, and then
showed to be due to Schottky anomaly. Motivated by this unsettled
issue we used IR and optical spectroscopy to directly test the
hypothesis of the HF state in Fe$_2$TiSn. Optical experiments are
perfectly suited for such a task because they probe both the
intra- and interband electronic excitations and have been
successfully employed in studies of $f$-electron HF systems
\cite{degiorgi99,dordevic01,wachter94,donovan97,sulewski88}.

The polycrystalline samples of Fe$_2$TiSn were grown by arc
melting under high purity argon on a copper hearth and have
previously been characterized by X-ray diffraction, electrical
resistivity, susceptibility, specific heat and XPS measurements
\cite{slebarski00}. For IR measurements, the samples were
mechanically polished until a mirror-like surface was achieved.
Near normal incidence reflectance $R(\omega)$ was measured at UCSD
in a broad frequency range 40-20,000 cm$^{-1}$ (approximately 5
meV - 2.5 eV) and temperature range (from 10 K to 300 K). To
obtain the absolute values of R($\omega$), the samples were coated
in-situ with gold or aluminum in the optical cryostat and the
spectrum of a metal-coated sample was used as a reference. This
procedure yields reliable absolute values of $R(\omega)$ and does
not require ambiguous corrections for diffuse reflectance
\cite{homes93}. The IR measurements were supplemented with
ultraviolet reflectance measurements up to 100,000 cm$^{-1}$ (12
eV) performed at room temperature at ETH.

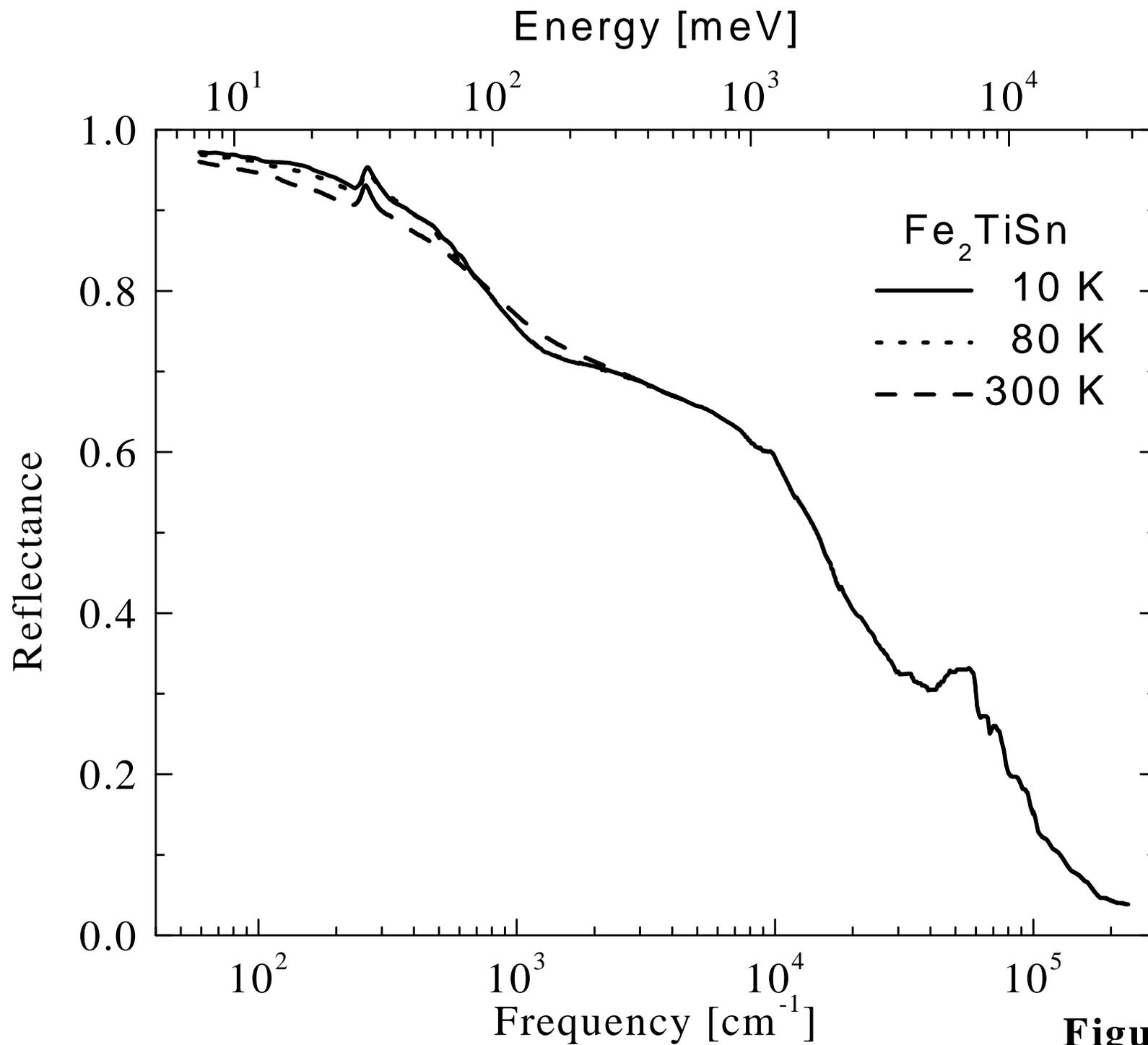
\begin{figure}
\caption{The reflectance data of Fe$_2$TiSn at 10, 80 and 300 K.
The spectra show anomalous temperature dependence in the
mid-infrared range.} \label{fig:ref}
\end{figure}

Figure \ref{fig:ref} shows the reflectance of Fe$_2$TiSn at
several selected temperatures. The general shape of reflectance is
metallic, but the exact position of the plasma minimum is obscured
because the free carrier response overlaps with interband
transitions. The far-infrared reflectivity ($\omega <$ 600
cm$^{-1}$) decreases as temperature increases, the behavior
typical for metallic systems. The mid-infrared reflectance,
however, shows more complicated temperature dependence: in the
range between 600 - 3,000 cm$^{-1}$, R($\omega$) decreases with
temperature. At higher frequencies ($\omega >$ 3,000 cm$^{-1}$),
the spectra are temperature independent. The peaks at $\sim$ 250
cm$^{-1}$ and $\sim$ 50,000 cm$^{-1}$ can be interpreted as an
optically active phonon and an interband transition, respectively.

The next step in data analysis is to perform a Kramers-Kronig (KK)
transformation on the raw reflectivity data in order to obtain the complex
optical conductivity $\sigma(\omega)= \sigma_{1}(\omega)+
i\sigma_{2}(\omega)$. For the low frequency extrapolation we used
a Hagen-Rubens formula, commonly employed for metals:
$R(\omega)=1-\sqrt{2 \omega \rho_{dc}/\pi}$, where $\rho_{dc}$ is
the dc resistivity. Several other extrapolations (such as a
straight line or $R(\omega)\sim\omega^2$) produced the same result
in the region where the data exist. A power law extrapolation
R($\omega$)$\sim \omega^{-4}$ was used for high frequencies.

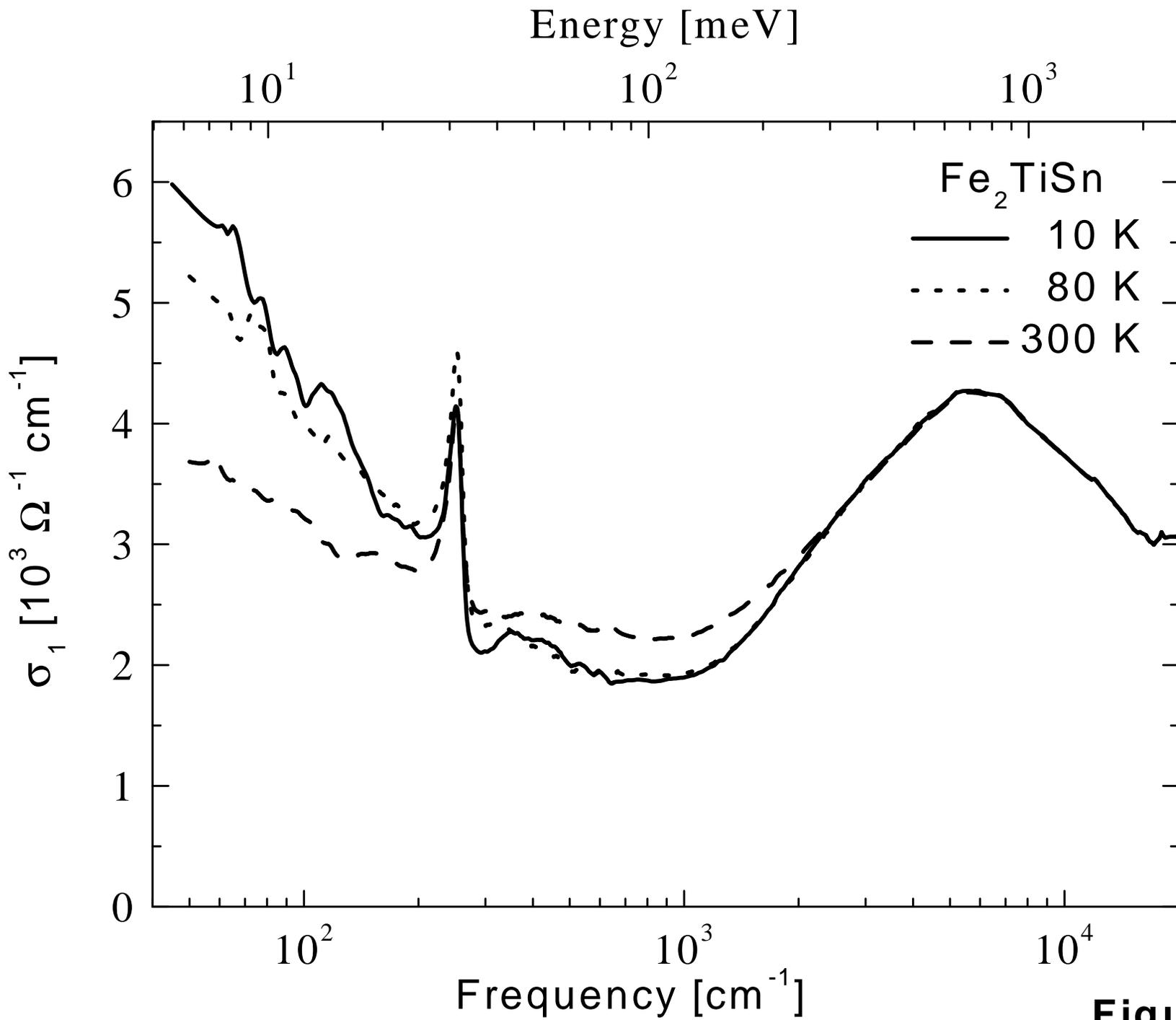
\begin{figure}
\caption{The optical conductivity $\sigma_1(\omega)$ of Fe$_2$TiSn
is characterized by a narrow Drude mode and an interband
transitions at around 7,000 cm$^{-1}$. A strong peak at 250
cm$^{-1}$ is an optically active phonon mode.} \label{fig:cond}
\end{figure}

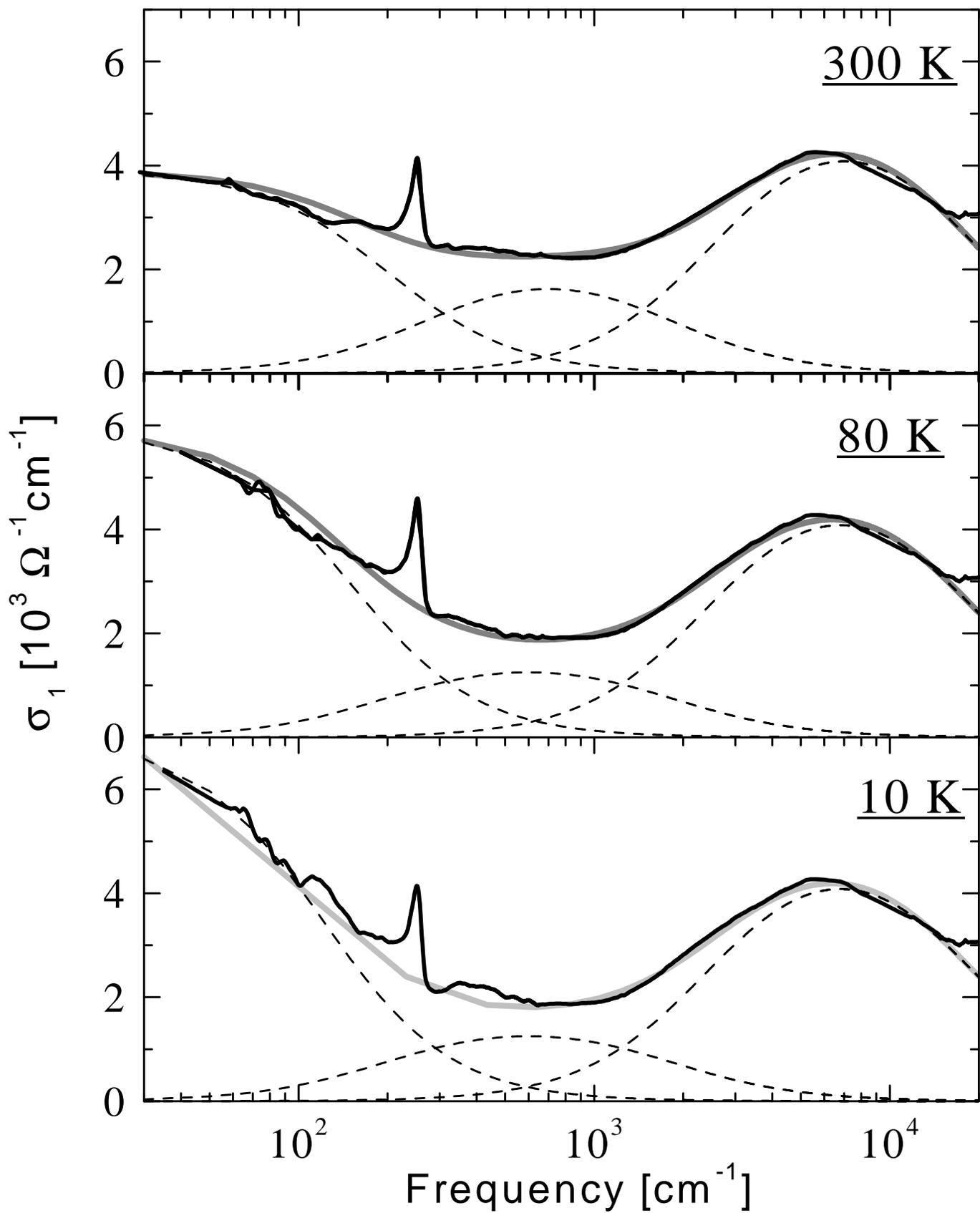
\begin{figure}
\caption{Drude-Lorentz fits of the optical conductivity
$\sigma_1(\omega)$ at 300, 80 and 10 K. As temperature decreases
the Drude peak narrows, but its spectral weight appears to be
conserved. The 7,000 cm$^{-1}$ peak is only weakly temperature
dependent.} \label{fig:fits}
\end{figure}

Figure \ref{fig:cond} shows the real part of the optical
conductivity $\sigma_1(\omega)$.
The spectra of Fe$_2$TiSn are characterized by a Drude-like mode
with a width of about 120 cm$^{-1}$ at 10 K and a broad peak
centered around 7,000 cm$^{-1}$ (0.87 eV). As temperature
increases, the zero energy peak broadens, whereas the peak at
7,000 cm$^{-1}$ displays almost no T-dependence. In order to
quantify these changes we first employ a conventional
Drude-Lorentz model. Fits including a Drude mode and a $single$
Lorentzian at 7,000 cm$^{-1}$ failed to produce satisfactory
results. We succeeded in accurately reproducing the
$\omega$-dependence of $\sigma_1(\omega)$ at all temperatures with
a set of $two$ Lorentz oscillators in addition to the Drude term
\cite{phononfit}:

\begin{equation}
\sigma(\omega)=\frac{1}{4 \pi}\frac{\omega_p^2 \tau}{1-i \omega \tau}+
\frac{1}{4\pi}\sum_{j=1}^2\frac{i \omega \omega_{pj}^2}
{ \omega^{2} - \omega_{j}^{2} + i \gamma_{j}\omega}.
\label{eq:drude}
\end{equation}
The first term represents a Drude free-electron component, where
$\omega_p^2=4\pi n e^2/m^*$ is the plasma frequency ($n$ is the
carrier density and m$^*$ is the carrier effective mass) and
1/$\tau$ is the carrier scattering rate. The last two terms in
Eq.~\ref{eq:drude} are the Lorentzian oscillators centered at
$\omega_j$, with the width $\gamma_j$ and plasma frequency
$\omega_{pj}$. The best fits for 300, 80 and 10 K are shown in
Fig.~\ref{fig:fits} with gray lines; the three individual
components of Eq.~\ref{eq:drude} are shown with dashed lines.
Table \ref{heusler} summarizes the fitting parameters. We
emphasize here that these parameters are unique, since no other
values can reproduce the quality of the fits shown in
Fig.~\ref{fig:fits}.

As can be seen the scattering rate, 1/$\tau$ monotonically
decreases with decreasing temperature, whereas the plasma
frequency of the Drude component is essentially temperature
independent ($\omega_p\simeq$ 7,000 cm$^{-1}$). Such behavior is
typical of conventional metals \cite{wooten,fawcett88,dordevic01b}
and provides a strong argument against a HF state in Fe$_2$TiSn.
Indeed the hallmark of the latter state is a strong increase of
m$^*$ below a temperature T$^*$ characteristic of a given
material. This is equivalent to a drastic reduction of the
oscillator strength of the Drude component ($\omega_p^2\sim
1/m^*$) (Ref.~\onlinecite{degiorgi99,dordevic01}),
which is not observed in our data.

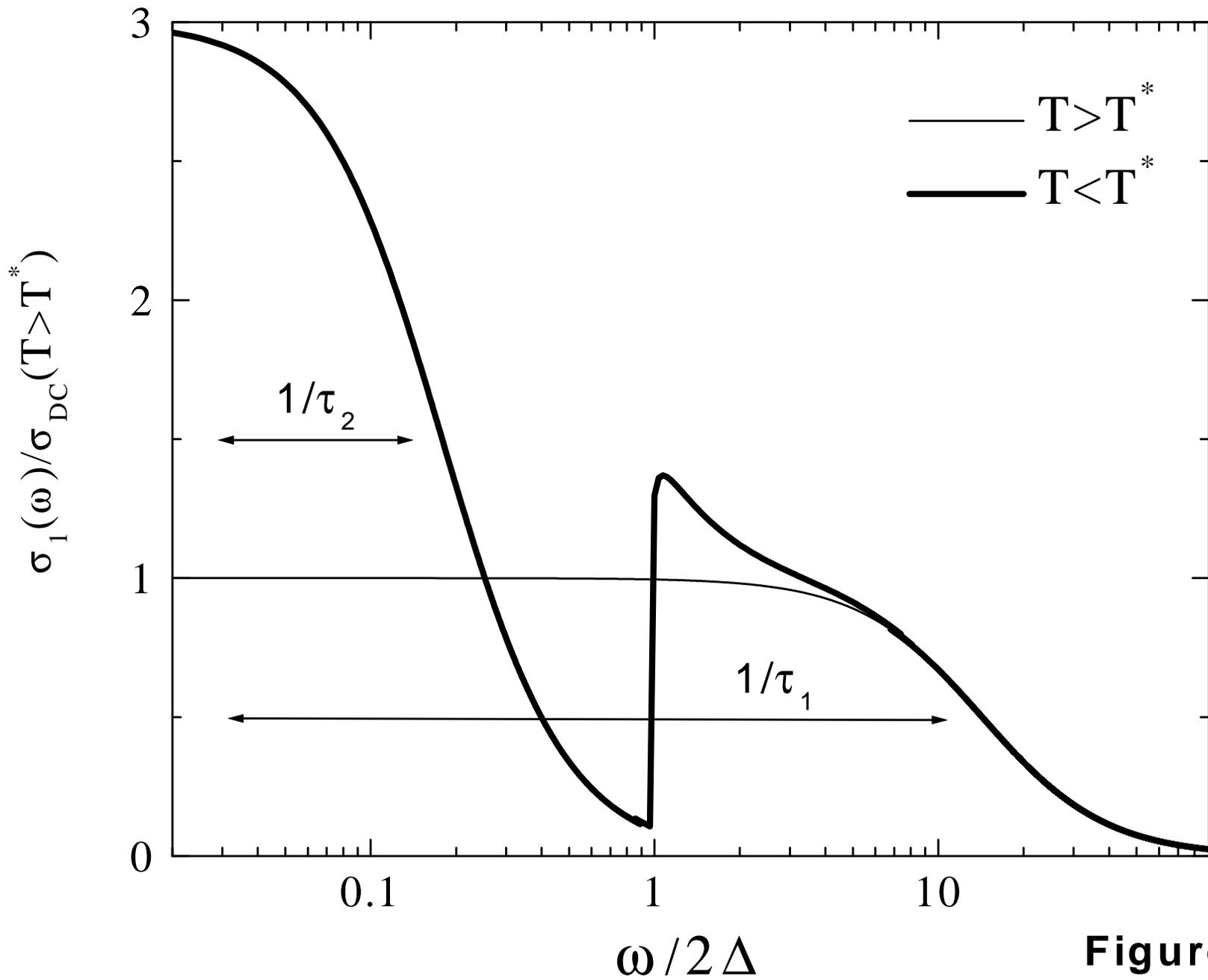
\begin{figure}
\caption{Schematic behavior of the frequency dependent optical
conductivity $\sigma_1(\omega)$ in an $f$-electron HF system. Note
that below T$^*$ both 1/$\tau$ and $\omega_p$ of the Drude mode
are strongly reduced.}
\label{fig:hf}
\end{figure}

Additional evidence against a HF state in Fe$_2$TiSn comes from
the line-shape analysis of the $\sigma_1(\omega,T)$ spectra. The
HF behavior has been shown before to leave characteristic
fingerprints in the optical spectra of such systems
(Ref.~\onlinecite{degiorgi99,dordevic01,degiorgi01}). Within a so
called "hybridization scenario", hybridization between free
carriers and localized $f$-electrons leads to a gap in the density
of states that develops below the temperature T$^*$. Excitations
across this hybridization gap, in addition to intraband
absorption, give rise to the optical conductivity
$\sigma_1(\omega)$ schematically shown in Fig.~\ref{fig:hf}. At
high temperatures (T$>$T$^*$), the conductivity of many HF systems
follows a simple Drude response
\cite{degiorgi99,dordevic01,degiorgi01} (thin solid line in
Fig.~\ref{fig:hf}). At T$<$T$^*$ two processes occur
simultaneously: 1) the width of the Drude mode is collapsing so
that $1/\tau_2 \ll 1/\tau_1$, and 2) a finite frequency peak due
to excitations across the hybridization gap appears. The latter
contribution to $\sigma_1(\omega)$ shown in Fig.~\ref{fig:hf} was
calculated within a BCS model with coherence factors type I
(Ref.~\onlinecite{dordevic01,gapcomm}).

The $\sigma_1(\omega)$ data of Fe$_2$TiSn (Fig.~\ref{fig:fits})
look qualitatively similar to the model spectra of
Fig.~\ref{fig:hf}: there is a Drude mode that narrows with
temperature and a finite frequency peak at around 700 cm$^{-1}$.
However, it appears that the temperature dependence of the latter
excitation is completely uncorrelated with the Drude mode. Unlike
HF systems, the spectral weight of the finite frequency peak
$\omega_{p1}$ $decreases$ slightly when lowering the temperature
(Table \ref{heusler}). Therefore, based on these findings one
cannot interpret the 700 cm$^{-1}$ peaks as being due to
excitations across the hybridization gap. Instead we speculate
that it is a low lying interband transition. A similar feature has
not been seen in Fe$_2$VAl (Ref.~\onlinecite{okamura00}); the data
for Co$_2$TiSn does not extend low enough \cite{shreder00}.

The absolute value of the Drude plasma frequency $\omega_p$
implies carrier density as small as n$\sim 5 \times 10^{20}$
cm$^{-3}$, under the assumption $m^*=m_0$ ($m_0$ is a free
electron mass). This value of $n$ is very similar to that found in
both Co$_2$TiSn (Ref.~\onlinecite{shreder00}) and Fe$_2$VAl
(Ref.~\onlinecite{okamura00,feng01}). It is striking that even
though the carrier density is essentially the same in Fe$_2$TiSn
and Fe$_2$VAl, the former compound shows metallic, whereas the
latter displays an insulating temperature dependence of the dc
resistivity. This indicates that localization (probably due to
disorder) plays an important role in the charge dynamics. For
Fe$_2$TiSn, we estimate the carrier mean free path $l\simeq$ 80
$\AA$ at room temperature. Such a long mean free path signals
that, unlike Fe$_2$VAl (Ref.~\onlinecite{feng01}), the Boltzmann
formalism should still hold for Fe$_2$TiSn.

In addition to the free electron Drude component and a low-lying
transition at 700 cm$^{-1}$, the optical spectrum of Fe$_2$TiSn
exhibits another excitation at around 7,000 cm$^{-1}$
(Fig.~\ref{fig:fits} and Tab.~\ref{heusler}). A similar peak at
7,000 cm$^{-1}$ has been seen before in Fe$_2$VAl
(Ref.~\onlinecite{okamura00}). This is not unexpected, however, as
the band structure calculations for both compounds show that the
bands around the Fermi level are predominantly coming from Fe
$d$-orbitals \cite{weht98,singh98,weinert98,guo98,slebarski00}. In
Fe$_2$VAl, the 7,000 cm$^{-1}$ peak has been interpreted as being
due to excitations across a pseudogap \cite{okamura00}. The band
structure calculations for Fe$_2$TiSn
(Ref.~\onlinecite{slebarski00,jezirski01}) have also predicted the
existence of a partial gap (pseudogap) at the Fermi level of about
0.5 eV. Although the position of the peak in Fe$_2$TiSn does not
exactly agree with this predicted value, we believe that it can be
interpreted as being due to excitations across such a gap. Namely
the magnitude of the gap has been shown before to depend strongly
on atomic disorder \cite{jezirski01}, which is difficult to
control during sample growth.

In conclusion, our IR results do not support the notion that the
Kondo interaction plays a dominant role in the charge dynamics of
Fe$_2$TiSn. We find that the effective mass of free carriers is
essentially temperature independent. By analogy with Fe$_2$VAl, we
suggest that the apparent mass enhancement at low temperatures is
due to a Schottky anomaly arising from magnetic clusters. The
latter effect can also explain the anomalous temperature
dependence of specific heat and dc resistivity at low
temperatures. The free electron contribution to the optical
conductivity of Fe$_2$TiSn appears to be conventional Drude-like,
with a small carrier density and relatively long mean free path.
The interband transition at 7,000 cm$^{-1}$ can be interpreted as
excitations across a pseudogap predicted in band structure
calculations.

We thank E.J.~Singley for useful discussions. The research at UCSD
supported by DOE, NSF and the Research Corporation.

$^{**}$ Also at Institute of Physics, University of
Silesia, 40-007 Katowice, Poland

\begin{table}
\caption{\label{heusler}
Fitting parameters from Eq.~\ref{eq:drude}, with all the values
given in the units of $cm^{-1}$.}
\centering
\begin{tabular}{|r||cc|ccc|ccc|}
& $\omega_p$    & 1/$\tau$ & $\omega_{j1}$ & $\gamma_1$ & $\omega_{p1}$ &
$\omega_{j2}$ & $\gamma_2$ & $\omega_{p2}$ \\ \hline
 10 K & 7,100 &  120 & 600 & 2,000 & 12,250 & 6,800 & 21,000 & 71,700 \\
 80 K & 7,200 &  150 & 600 & 2,000 & 12,250 & 6,800 & 21,000 & 71,700 \\
300 K & 6,870 &  200 & 700 & 2,000 & 14,000 & 7,000 & 21,000 & 71,700 \\
\end{tabular}
\end{table}


\begin{references}
\bibitem{phononfit} We did not atempt to fit the phonon mode at
250 cm$^{-1}$.
\bibitem{gapcomm} We have performed a calculation of
[$\sigma_{1}(\omega,T<T^*)/\sigma_{1}(\omega,T>T^*)$], the ratio
of the optical conductivity below and above the coherence
temperature. The method follows closely the BCS calculation
\cite{tinkham96}, except that the coherence factors are of type I,
as opposed to type II in conventional superconductors.
\end{references}
\end{document}